\documentclass[aps,twocolumn,showpacs,preprintnumbers,prb]{revtex4}
\usepackage{graphicx}

\newcommand{\Ai}{\mathrm{Ai}}
\newcommand{\R}{{\bf r}}
\newcommand{\RR}{{\bf r}'}

\newcommand{\Q}{\mathbf{q}}
\newcommand{\UP}{n_{\uparrow}}
\newcommand{\DN}{n_{\downarrow}}

\newcommand{\be}{\begin{equation}}
\newcommand{\ee}{\end{equation}}
\newcommand{\bea}{\begin{eqnarray}}
\newcommand{\eea}{\end{eqnarray}}
\newcommand{\bean}{\begin{eqnarray*}}
\newcommand{\eean}{\end{eqnarray*}}

\newenvironment{acknowledgment}{{\flushleft \bf Acknowledgments:}}{}

\begin{document}

\title{ Exchange-Correlation Energy Functional Based on the Airy-Gas Reference System }
\author{Lucian A. Constantin, Adrienn Ruzsinszky and John P. Perdew}
\affiliation{ Department of Physics and
Quantum Theory Group, Tulane University, New Orleans, LA 70118}

\date{\today}

\begin{abstract}
In recent work, generalized gradient approximations (GGA's) have been constructed from the
energy density of the Airy gas for exchange  but not for correlation.
We report the random phase approximation (RPA) conventional correlation energy density of the 
Airy gas, the simplest edge electron gas, in which the auxiliary noninteracting electrons
experience a linear potential. 
By fitting the Airy-gas RPA exchange-correlation energy density and making an accurate short-range 
correction to RPA, we propose a simple beyond-RPA GGA 
density functional ("ARPA+") for the exchange-correlation energy. Our functional, tested for 
jellium 
surfaces, atoms, molecules and solids, improves 
mildly over the local spin density approximation
for atomization energies and lattice constants without 
much worsening the already-good surface exchange-correlation energies.
\end{abstract}


\maketitle

\section{Introduction}
\label{sec1}
\noindent

In Kohn-Sham density functional theory \cite{KS}, the ground-state density and energy 
of interacting electrons in a scalar external potential $v(\R)$ are computed efficiently via a 
selfconsistent calculation for an auxiliary system of noninteracting electrons in a scalar effective 
potential $v_{eff}(\R)$.
Once the exchange-correlation energy as a functional of the electron density has been approximated, 
its
functional derivative provides the exchange-correlation contribution to $v_{eff}(\R)$.  
By itself, the deviation of $v_{eff}(\R)$
from the constant chemical potential determines the electron density and thus the correlation
energy.
Typical approximations
are designed to be exact for a reference system, most often the uniform electron gas in which the 
auxiliary noninteracting electrons see a constant or uniform $v_{eff}$.  Sometimes additional exact 
constraints or fits to experiment are also built into the approximation.  Recently Kohn and Mattsson 
\cite{KM1} have
proposed as a more realistic reference system the edge electron gas, in which $v_{eff}(\R)$ 
varies more
or less linearly near the edge surface of the density.  While the uniform gas could be (and is) a 
good
reference for a bulk solid, the edge electron gas could be at least as good for a bulk solid and 
better for
solid surfaces, molecules, and atoms, which have regions where the electron density evanesces.

The edge surface of any electron system is defined \cite{KM1} by $v_{eff}(\R)=\mu$, where 
$v_{eff}(\R)$ is the exact 
Kohn-Sham \cite{KS} (KS) effective potential and $\mu$ is the chemical potential.
Outside this classical turning surface,
all noninteracting electrons tunnel into a barrier.
The simplest example of an edge electron gas is the Airy gas, where any electron feels
a linear
effective potential \cite{KM1}, and thus the normalized one-particle eigenfunctions are
proportional to the Airy function.
The Airy gas has not only a
surface-like region, but also a region of high and slowly-varying (Thomas-Fermi-like) electron
density where the local density approximation (with uniform-gas input) is accurate 
\cite{KM1,PCSB} 
for the noninteracting kinetic, exchange, and correlation energy densities.

The Airy gas has appeared before in density functional theory:
(1) The effective finite-linear-potential model gives remarkably good
results for
the jellium surface problem, where the orbitals of this model are approximated with plane waves 
inside the 
bulk, Airy functions near the surface, and  exponential functions far in the 
vacuum \cite{SMF,SS3,SB}. (2) 
Baltin \cite{Ba} constructed a generalized gradient approximation
(GGA) for the orbital kinetic energy from the Airy-gas kinetic energy density, but his 
approximation 
does not recover the second-order gradient expansion for the kinetic energy density 
\cite{Ki,BJC} and 
is poor for atoms and molecules \cite{GB,Vi1}.  However, the kinetic energy density of the Airy 
gas 
\cite{Vi1} can still be a starting point for construction of GGA kinetic energy functionals 
that can 
be more
accurate for atoms, molecules, jellium clusters, and jellium surfaces \cite{Vi1,CR}.  The trick 
is to fit a
GGA \emph{plus a $\nabla^2 n$ term integrating to zero} to the Airy-gas kinetic energy density.
 
The exchange energy density of the Airy gas \cite{KM1} was fitted \cite{Vi2,AM05} with a 
function 
dependent on the density and its gradient. Thus, Vitos \emph{et. al} \cite{Vi2} developed a 
GGA exchange energy functional (LAG or local Airy-gas GGA) that was used with the local spin-density 
approximation (LSDA) correlation energy. This exchange-correlation (xc) energy functional gives 
results for atoms very close to, but better than, the LSDA ones, and its accuracy for 
atomization energy of diatomic molecules is similar to that of the PBE GGA \cite{PBE},  
while for bulk systems the results of LAG GGA are close to the PBEsol GGA \cite{PRCVSCZB} and 
to experimental values. 
However, the jellium  xc surface energies of LAG are far too low (lower even than those of the PBE 
GGA).
Armiento and Mattsson \cite{AM05,AM05b} proposed an xc energy functional 
(AM05 GGA) 
using a better fit for the Airy gas exchange energy density and a correlation energy
functional constructed such that
the AM05 xc jellium surface energies fit the RPA+ \cite{YPK} values (RPA plus a GGA short-range 
correction).
AM05 is also based on the subsystem functional approach \cite{AM11}, which permits an 
interpolation
between a uniform-gas reference for the bulk of a solid and an Airy-gas reference for the surface.
(Since the Airy-gas reference system
by  itself provides such an interpolation, we make no further interpolation here.)
AM05 slightly improves the accuracy of LAG GGA for bulk systems.

Because the correlation energy density of the Airy gas was unknown, the LAG GGA and
AM05 GGA used in their construction only the Airy-gas exchange energy density.
In this paper we compute the correlation 
energy density of the Airy gas in the random phase approximation (RPA), and fit it
to a GGA (ARPA). 
As in Refs. \cite{Vi2} and \cite{AM05}, our fit is made without
regard to exact constraints on $E_{xc}[\UP,\DN]$.
The Airy gas is a system of 
delocalized electrons where the self-interaction correction has no effect, 
and where the GGA correction \cite{YPK} to the integrated RPA energy should be accurate.
Our functional, including this GGA correction to RPA, will be called ARPA+.

Unlike energies, energy densities of non-uniform systems are not unique.  It is not clear to us 
that the
conventional choice
for the exchange-correlation energy density (made in Refs. \cite{Vi2,AM05}, and here) is 
optimal.  It is not our intention
here to either endorse or criticize this choice, but simply to see 
what GGA is obtained from the Airy-gas reference system
within a consistent implementation for correlation as well as exchange.

AM05, PBEsol, and ARPA+ are of special interest as candidates for a "GGA for solids"
providing better lattice constants and surface energies than standard GGA's like PBE, possibly
at the cost of a worsened description of atoms and molecules. There have been several recent
articles commenting on or testing for solids the LAG, AM05, and PBEsol GGA's 
\cite{MAM,PRCVSCZB2,MWA,RKV,CPRPLPVA}.

Our paper is organized as follows. In section \ref{sec2}, we propose a simple model for the 
Airy gas. In section \ref{sec3}, we construct the ARPA+ GGA xc energy functional from our 
Airy gas model. In section \ref{sec4} we test the ARPA+ GGA for atoms, molecules, jellium 
surfaces and bulk solids. In section \ref{sec5}, we summarize our conclusions. 

\section{The Airy gas model}
\label{sec2}
\noindent

The simplest example of an edge electron gas is the Airy gas that is translationally invariant in 
the plane of the surface ($z=0$) and has 
the effective potential \cite{KM1,noteVE} 
\begin{equation}
v_{eff}(z)=\left\{ \begin{array}{lll}
-F z,     & -\infty < z < L \;\;\;\;(F>0)\\
\infty,     & z\geq L \;\;\;\;(L/l\rightarrow\infty).\\
                                    \end{array}
\right.
\label{e1}
\end{equation}
Here $F=|d v_{eff}(z)/dz|$ is the slope of the effective potential and the characteristic 
length 
scale 
\begin{equation}
l=(2F)^{-1/3}
\label{e2}
\end{equation} 
is approximately the edge region thickness \cite{KM1}.
(Unless otherwise stated, atomic units are used throughout,
i.e., $e^2=\hbar=m_e=1$.)

The KS orbitals are 
$\Psi_{j,\bf{k}_{||}}(\R)=\phi_j(z)\frac{1}{\sqrt{A}}e^{i\bf{k}_{||}\R_{||}}$, where
$\bf{k}_{||}$  and $\R_{||}$ are the wavevector and the position vector parallel to the 
plane of the surface, $A$ is the cross-sectional area, and the orthonormal eigenfunctions 
$\phi_j(z)$ satisfy the equation
\begin{equation}
(-\frac{1}{2}\frac{d^2}{dz^2}-Fz-\epsilon_j)\phi_j(z)=0,
\label{e3}
\end{equation}
with the boundary conditions
\begin{equation}
\phi_j(-\infty)=\phi_j(L)=0.
\label{e4}
\end{equation}
They are given by the Airy functions 
\begin{equation}
\phi_j(z)=a\Ai(-\frac{z}{l}-\frac{\epsilon_j}{\epsilon}),
\label{e5}
\end{equation}
where $\epsilon=(F^2/2)^{1/3}$ is the Airy gas characteristic energy scale, $a$ is the 
normalization constant, and $\epsilon_j$ is the $j$-th eigenvalue calculated from the boundary 
condition $\phi_j(L)=0$.
The Airy gas density is 
\begin{equation}
n(z)=\sum_j^{occ}\phi_j^2(z)|\epsilon_j|/\pi.
\label{e6}
\end{equation}
We recall that all 3D states with energy up to $\mu=0$ are occupied.
Thus the Airy gas is completely determined by the length $l$ and the energy $\epsilon$.

In the limit $L/l\rightarrow\infty$, the normalization constant is \cite{KM1}
\begin{equation}
a=\frac{\pi^{1/2}}{(Ll)^{1/4}},
\label{e7}
\end{equation}
and the eigenvalues are \cite{KM1}
\begin{equation}
\epsilon_j=-j(\frac{l}{L})^{1/2}\pi\epsilon.
\label{e8}
\end{equation}
So, the density of the Airy gas is
\begin{equation}
n(z)=l^{-3}n_0(\eta),\;\;\;\;\;\eta=z/l,
\label{e9}
\end{equation}
where
\begin{equation}
n_0(\eta)=\frac{1}{2\pi}\int^\infty_0 \Ai^2(\eta'-\eta)\eta'd\eta'.
\label{e10}
\end{equation}

Let us consider a model for the Airy gas that is described by Eqs. (\ref{e1}) - (\ref{e6}), 
but instead of choosing $L/l\rightarrow\infty$ we take $L/l=20$
for computational convenience.
Such a system has 19 
occupied orbitals $\phi_j(z)$ and can accurately describe the Airy gas. The normalization 
constants of Eq. 
(\ref{e5}) and the eigenvalues $\epsilon_j$ are computed numerically. 
Such an approach is similar to jellium slabs that are described by a finite number of occupied 
orbitals in the $z$-direction and that can accurately predict the surface energies of  
semi-infinite jellium surfaces \cite{PE2}. 

\begin{figure}
\includegraphics[width=\columnwidth]{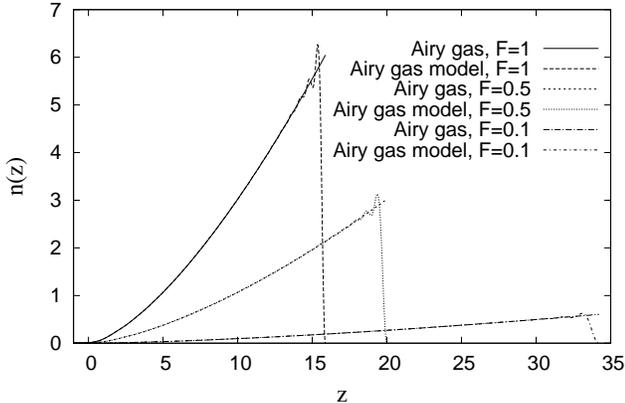}
\caption{ Electron density (electrons/bohr$^3$) of the Airy gas and of our model versus $z$ 
(bohr), for 
several slopes of 
the effective potential ($F=0.1$ making $l=1.710$, $F=0.5$ making $l=1.000$,
$F=1$ making $l=0.793$). The edge is 
at $z=0$. 
}
\label{f1}
\end{figure}

We select three values $F=0.1$, $F=0.5$, and $F=1$ for the slope of the effective potential. The 
accuracy of the model does not depend on the $F$ value.
In Fig. \ref{f1} we show the densities of the Airy gas and of our Airy gas model for the 
chosen values of the slope $F$. We see the exact Airy gas densities and the modeled ones can 
not be distinguished until $z\sim L=20\cdot l$ where the densities of our model have 
oscillations until they vanish.
%

\begin{figure}
\includegraphics[width=\columnwidth]{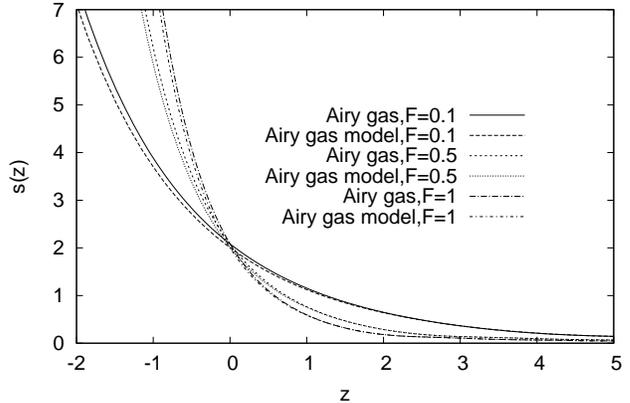}
\caption{ Reduced gradient $s(z)$ versus z, of the Airy gas and our model, for several 
slopes of 
the effective potential ($F=0.1$, 0.5, and 
1.) The edge is at $z=0$.}
\label{f2}
\end{figure}

Important ingredients of any GGA functional are the density $n(\R)$ and the reduced
density gradient
\begin{equation}
s(\R)=|\nabla n(\R)|/[2k_F(\R)n(\R)],
\label{e11}
\end{equation}
where $k_F(\R)=(3\pi^2 n(\R))^{1/3}$ is the Fermi wavevector. (The dimensionless density
gradient $s(\R)$ measures the variation of the density over a Fermi wavelength
$\lambda_F=2\pi/k_F$.) In Fig. \ref{f2} we compare the reduced gradients of our model and of 
the exact Airy gas. Up to $s=2$, the model nicely matches the exact Airy gas, and it is accurate 
for any value of $s$. (We note that $s$ values bigger than 3 are
found in the tail of an atom or molecule, where the electron density is negligible. We also 
note that 
in most bulk solids the maximum \cite{CPRPLPVA}
value of the reduced gradient is smaller than 
2.)   
%
%
Figs. \ref{f1} and \ref{f2} demonstrate that our model is accurate, and thus we can use it for 
the calculation of the Airy gas correlation energy.

\section{ RPA correlation energy density of the Airy gas, and the construction of the ARPA+ GGA}
\label{sec3}
\noindent

The conventional xc energy density at a point is $n\epsilon_{xc}$, where $n$ is the local 
electron density 
and 
$\epsilon_{xc}$ is
the conventional xc energy per particle.
Let us consider the spin-unpolarized Airy gas model with the edge plane at $z=0$. Using its
translational invariance in a plane perpendicular to the $z$ axis, and the so-called 
adiabatic-connection fluctuation-dissipation
theorem \cite{LP,GL,HG,PE2} (ACFDT), the exact expression for the conventional xc energy per 
particle
at point $z$ is \cite{LP,GL,PE2}
\begin{eqnarray}
\epsilon_{xc}(z)=\frac{1}{2}\int \frac{d \Q_{||}}{(2\pi)^2}\int
d\acute{z}\; v(z,\acute{z},q_{||})[-\frac{1}{\pi n(z)}\nonumber\\
\times \int^1_0 d\lambda\int^\infty_0 d\omega\chi^\lambda
(z,\acute{z};q_{||},i\omega)-\delta(z-\acute{z})],
\label{e12}
\end{eqnarray}
where $\Q_{||}$ is the wavevector parallel to the surface, and
$\chi^\lambda$ and $v$ are the two-dimensional Fourier transforms of
 the interacting density response
function at the coupling strength $\lambda$ and of the Coulomb potential
respectively.
The substitution of $\chi^\lambda$ with the non-interacting density response function $\chi^0$ into 
Eq.(\ref{e12})
 yields the exact $\epsilon_x(z)$
(expressible in terms of occupied orbitals only, although $\chi^0$
requires also the unoccupied orbitals).
The density response function
obeys the screening integral Dyson-like equation \cite{GDP}
\begin{eqnarray}
\chi^\lambda(\R,\RR,\omega)=\chi^0(\R,\RR,\omega)+\int d\R_1 d\R_2
\chi^0(\R,\R_1,\omega) \nonumber\\
\times\{v^\lambda(\R_1,\R_2)+f^\lambda_{xc}[n](\R_1,\R_2,\omega)\}
\chi^\lambda(\R_2,\RR,\omega),
\label{e13}
\end{eqnarray}
where $v^\lambda(\R_1,\R_2)=\lambda/|\R_1-\R_2|$ and
$f^\lambda_{xc}[n](\R_1,\R_2,\omega)=\delta v^\lambda_{xc}[n]
(\R_1,\omega)/\delta n(\R_2,\omega)$ is the exact xc kernel.
Here $v^\lambda_{xc}[n]$ is the exact frequency-dependent xc 
potential at coupling strength $\lambda$.
 Obviously, the exact xc kernel is unknown and it has to be approximated.
Approximations of the xc kernel are usually constructed from the uniform 
electron gas \cite{CP,PP,JGGDG}, 
and have not been tested sufficiently for  nonuniform systems.  
When
$f^{\lambda}_{xc}[n]({\bf r},{\bf r}';\omega)$ is taken to be zero,
Eq.~(\ref{e13}) reduces to the RPA. The RPA xc hole density is exact at 
large interelectronic separations such that it can correctly describe the xc hole density
of an electron far outside of a jellium surface \cite{CP2}, and its on-top hole is 
finite and well described by the LSDA-RPA \cite{YPK} on-top hole in the case of a jellium 
surface \cite{CP2}.  

Eqs. (\ref{e12}) and (\ref{e13}) can be generalized \cite{DLV} for systems with any relative 
spin 
polarization
\begin{equation}
\zeta=\frac{\UP-\DN}{\UP+\DN},
\label{e14}
\end{equation}
where $\UP$ and $\DN$ are the spin densities, $\UP+\DN=n$.
Thus for the Airy gas model, we choose to calculate the RPA correlation energy per particle at 
point z, from Eqs. (\ref{e12}) and (\ref{e13}), and to add the RPA+ short-range correction:
\begin{equation}
E^{RPA+}_{xc}=E^{RPA}_{xc}+(E^{GGA}_{xc}-E^{GGA-RPA}_{xc}),
\label{e15}
\end{equation}
where $E^{GGA}_{xc}$ is the PBE GGA \cite{PBE} xc energy, and $E^{GGA-RPA}_{xc}$ is the PBE-RPA 
GGA xc energy \cite{YPK}. 
The exchange contribution and the long-range correlation contribution cancel out of the bracketed 
term in Eq. (15), leaving only short-range correlation.
Because the self-interaction correction is not important for the Airy 
gas, Eq. (\ref{e15}) will give nearly the exact correlation energy of the Airy gas.

For the numerical evaluation of Eqs.~(\ref{e12}) and (\ref{e13}), we follow the method 
described in
Refs.~\cite{PE2} and \cite{eguiluz}, but instead of using the double- and single-cosine 
representations of the 
density response function and the density respectively, we use a grid on the $z$-axis for 
$\chi^\lambda(z,\acute{z};q_{||},i\omega)$ and $n(z)$. We find that the first 50 unoccupied 
orbitals $\phi_j(z)$ are enough for an accurate calculation. (Our grid on the $z$-axis can accurately 
describe the occupied and the first 50 unoccupied orbitals \cite{noteG}). 

The exchange energy for a spin-polarized system may be evaluated
from the spin-unpolarized version using the spin-scaling relation \cite{OP}:
\begin{equation}
E_{x}[n_{\uparrow} ,n_{\downarrow}]=\frac{1}{2}\{E_{x}[2 n_{\uparrow}]+E_{x}[2 
n_{\downarrow}]\},
\label{e16}
\end{equation}
and thus we only need to consider the spin-unpolarized case.
We fit the exchange energy per particle of the Airy gas model, using the non-linear 
least-square Levenberg-Marquardt method \cite{PTVF}, with the following 
expression
\begin{equation}
\epsilon_x^{A}(n(\R))=\epsilon_x^{LSDA}(n(\R))F^A_x(s(\R)),
\label{e17}
\end{equation}
where $\epsilon_x^{LSDA}=-3k_F/4\pi$ and the enhancement factor is
\begin{equation}
F^A_x=\frac{a_1s^{a_2}}{(1+a_3s^{a_2})^{a_4}}+\frac{1-a_5s^{a_6}+a_7s^{a_8}}{1+a_9s^{a_{10}}}
\label{e18}
\end{equation}
%
where $a_1=0.041106$, $a_2=2.626712$, $a_3=0.092070$, $a_4=0.657946$ are the parameters 
found in Ref.\cite{Vi2}, and $a_5=133.983631$, $a_6=3.217063$, $a_7=136.707378$, 
$a_8=3.223476$, $a_9=2.675484$, 
$a_{10}=3.473804$ are parameters found from our fitting procedure.
Eq. (\ref{e17}) recovers the correct LSDA for the uniform 
electron gas, and fits well the Airy gas exchange energy per particle for $s\leq 20$. ( Values 
of $s$ bigger than 20 are found only when the density is 
negligible. We recall that LAA of Ref. \cite{AM05} is a better fit than LAG or $\epsilon^A_x$
far outside the edge.)  
%
\begin{figure}
\includegraphics[width=\columnwidth]{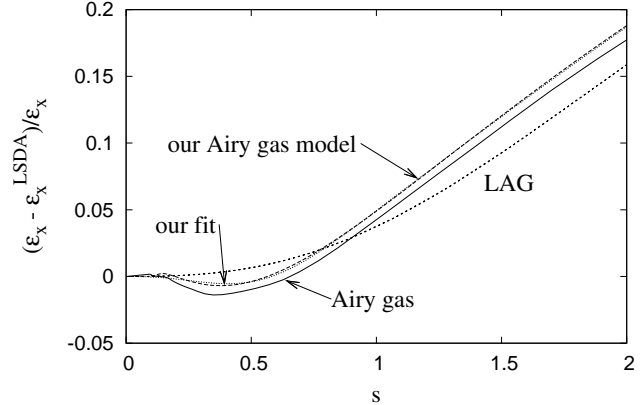}
\caption{$(\epsilon_{x}-\epsilon_{x}^{LSDA})/\epsilon_{x}$ versus the reduced gradient $s$ for
the Airy gas model, the Airy gas, the LAG GGA \cite{Vi2}, and for our fit (see Eqs. 
(\ref{e17}) and (\ref{e18})). 
The "Airy gas" curve uses $\epsilon_x$ and $\epsilon^{LSDA}_x$ of the Airy gas, whereas the 
other curves use $\epsilon_x$ and $\epsilon^{LSDA}_x$ of our model for the Airy gas.
The AM05 GGA \cite{AM05}, not shown in the figure, has the same behavior as 
the LAG GGA.}
\label{f3}
\end{figure}
%

In Fig. \ref{f3} we show $(\epsilon_{x}-\epsilon_{x}^{LSDA})/\epsilon_{x}$ versus the 
reduced gradient $s$ for several approximations. The Airy gas curve, as well as our Airy gas 
model curve, have a negative region around $s\approx 0.5$ that was not taken into account by 
the LAG GGA and AM05 GGA. 
We find this fine feature only
because we plot $(\epsilon_x-\epsilon_x^{LSDA})/\epsilon_x$ instead of $\epsilon_x$.
(This feature can also be seen in the inset of Fig. 1 of Ref. \cite{AM05}, but it was not 
taken 
into account in the construction of AM05.)
The second term of the right-hand-side of Eq. (\ref{e18})
models the exact behavior at small reduced gradients, whereas the first term of the 
right-hand-side of Eq. (\ref{e18}) has the same form as the parametrization proposed in Ref. 
\cite{Vi2}. We observe that our fit (Eqs. 
(\ref{e17}) and (\ref{e18})) is very close to the exact Airy gas model as well as to the exact 
Airy gas exchange energy per particle. 

We fit the RPA correlation energy per particle of the Airy gas of any spin polarization with 
the following expression, using again the non-linear
least-square Levenberg-Marquardt method \cite{PTVF} 
\begin{equation}
\epsilon_c^{ARPA}(r_s,\zeta,s_c)=\epsilon_c^{LSDA-RPA}(r_s,\zeta)F_c(s_c),
\label{e19}
\end{equation}
where $r_s$ is the local Wigner-Seitz radius [$n=3/(4\pi r_s^3)=k^3_F/3\pi^2$], $\zeta$ is the 
relative spin polarization of Eq. (\ref{e14}), $\epsilon_c^{LSDA-RPA}$ is the RPA correlation 
energy per particle of the uniform electron gas (see Ref. \cite{PW1}), and
\begin{equation}
s_c(\R)=\phi|\nabla n(\R)|/[2(3\pi^2)^{1/3}n(\R)^{7.9/6}],
\label{e20}
\end{equation}
with $\phi=[(1+\zeta)^{2/3}+(1-\zeta)^{2/3}]/2$ being a spin-scaling factor. The correlation 
enhancement factor is 
\begin{equation}
F_c=\frac{1+b_1s_c^3+b_2s_c^4}{1+b_3s_c^3+b_4s_c^4}
\label{e21}
\end{equation}
with $b_1=1.01453936$, $b_2=0.3255243$, $b_3=0.941597104$, and $b_4=0.587664306$.
Eq. (\ref{e21}) is a simple Pad\'{e} approximation that recovers the RPA behavior of the 
uniform electron gas when $s_c=0$. All the 
parameters were found by the fitting procedure, and not by 
constraints on the integrated correlation energy (which would suggest \cite{PBE}
an exponent of 7/6 and the appearance of $\phi$ in the denominator of Eq. (\ref{e20}), and a 
quadratic 
term in the
small-gradient expansion of Eq. (\ref{e21})). 
The irrelevance of some standard constraints may be related to the
absence \cite{PCR} of a second-order gradient expansion for the conventional correlation energy 
density.
Given $F$, $\epsilon^{RPA}_c$ is a function of $z$, and $s_c$ is a monotonic (hence invertible) 
function of $z$, 
so $\epsilon^{RPA}_c$ can be expressed as a function of $s_c$.
Since there is a one-to-one
correspondence between the $\epsilon_c^{RPA}$
and our $\epsilon_c^{ARPA}$, we can do the fitting.
The fitting was done for $s_c$ between 0 and 20.
%
\begin{figure}
\includegraphics[width=\columnwidth]{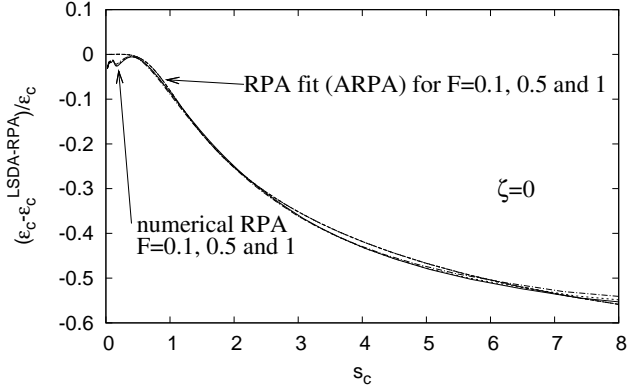}
\caption{$(\epsilon_{c}-\epsilon_{c}^{LSDA-RPA})/\epsilon_{c}$ of the spin-unpolarized 
($\zeta=0$) Airy gas model versus $s_c$ (see Eq. (\ref{e20})) for numerical RPA and 
our fit ARPA of Eq. (\ref{e19}), for several slopes of the effective potential 
($F=0.1$, 0.5, and 1).
Note that the numerical RPA has errors of order 2\% in the region of small reduced
gradient $s_c$.
}
\label{f4}
\end{figure}
%
\begin{figure}
\includegraphics[width=\columnwidth]{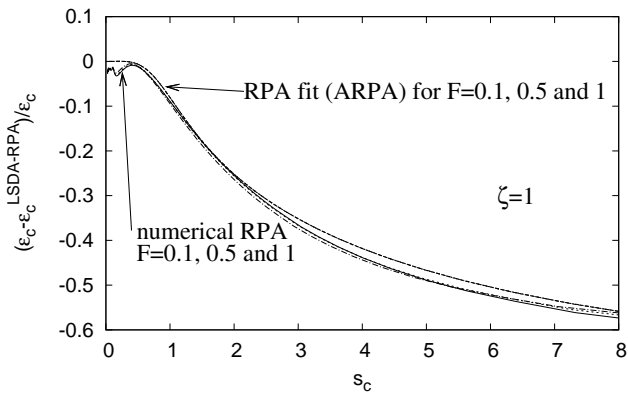}
\caption{$(\epsilon_{c}-\epsilon_{c}^{LSDA-RPA})/\epsilon_{c}$ of the fully-spin-polarized
($\zeta=1$) Airy gas model versus 
$s_c$ for numerical 
RPA and
our fit ARPA of Eq. (\ref{e19}), for several slopes of the effective potential 
($F=0.1$, 0.5, and 1).
Note that the numerical RPA has errors of order 2\% in the region of small reduced
gradient $s_c$.
}
\label{f5}
\end{figure}

In Figs. \ref{f4} and \ref{f5} we show $(\epsilon_{c}-\epsilon_{c}^{LSDA-RPA})/\epsilon_{c}$ 
versus
$s_c$ for the 
spin-unpolarized Airy gas model ($\zeta=0$) and fully-spin-polarized Airy gas model ($\zeta=1$)
respectively, for the slopes of the the effective
potential used in Figs. \ref{f1} and \ref{f2} ($F=0.1$, 0.5, and 1).
We note that our numerical calculation is accurate for $s_c\geq\sim 0.3$, see Ref. \cite{noteG}.
We see in both figures that the numerical RPA
correlation energy density 
does not depend much on the slope
value $F$ when they are plotted against
$s_c$, motivating our definition of $s_c$ in Eq. (\ref{e20}) and making the fit of
the RPA correlation energy per particle independent of the $F$ value \cite{noteSc} (see Eqs. 
(\ref{e19}) and 
(\ref{e21})).
For $s_c\leq 0.5$ the ARPA of Eq. (\ref{e19}) is close to exact even if 
it does not match well the detailed exact behavior, 
as it does in the region $0.5\leq s_c\leq 10$.

Overall we consider  
\begin{equation}
\epsilon_{xc}^{ARPA}=\epsilon_{x}^{A}+\epsilon_{c}^{ARPA}
\label{e21bis}
\end{equation}
an xc GGA functional that fits very well the Airy gas RPA xc energy density. Thus making the 
RPA+ 
short-range correction (see Eq. (\ref{e15})) to ARPA GGA, we propose the following GGA xc functional 
(ARPA+ 
GGA) constructed from the Airy gas
\begin{equation}
\epsilon_{xc}^{ARPA+}=\epsilon_{xc}^{ARPA}+(\epsilon_{c}^{PBE}-\epsilon_{c}^{PBE-RPA}).
\label{e22}
\end{equation}

The nonlocality of a GGA is displayed by the enhancement factor \cite{PEZB,PTSS}
\begin{equation}
F_{xc}^{GGA}=\frac{\epsilon_{xc}^{GGA}(\UP,\DN,\nabla\UP,\nabla\DN)}{\epsilon_x^{unif}(n)},
\label{e23}
\end{equation}
$\epsilon_x^{unif}(n)$ being the exchange energy per particle of a spin-unpolarized
uniform electron gas.
For a spin-unpolarized system in the high-density limit ($r_s\rightarrow 0$), the exchange
energy is dominant and
Eq. (\ref{e23}) defines
the exchange enhancement factor $F^{GGA}_x=\epsilon^{GGA}_x(n,\nabla
n)/\epsilon^{unif}_x(n)$.
%
\begin{figure}
\includegraphics[width=\columnwidth]{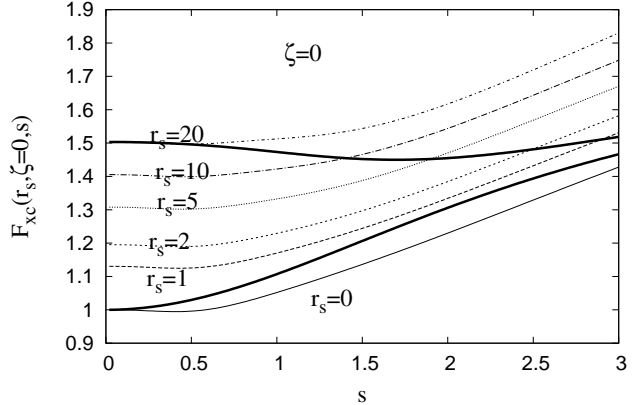}
\caption{ Enhancement factor $F_{xc}$ (see Eq. (\ref{e23}) for the spin-unpolarized case 
($\zeta=0$), as a
function of the reduced gradient $s$ for several values of $r_s$ ($r_s=0$, 1, 2, 5, 10, and 
20). The thin lines represent the 
ARPA+ enhancement factor whereas the thick lines are the
PBEsol enhancement factor for $r_s=0$ and $r_s=20$ respectively. The LSDA is 
$F_{xc}(r_s,\zeta=0,s=0)$.}  
\label{f6}
\end{figure}
%
\begin{figure}
\includegraphics[width=\columnwidth]{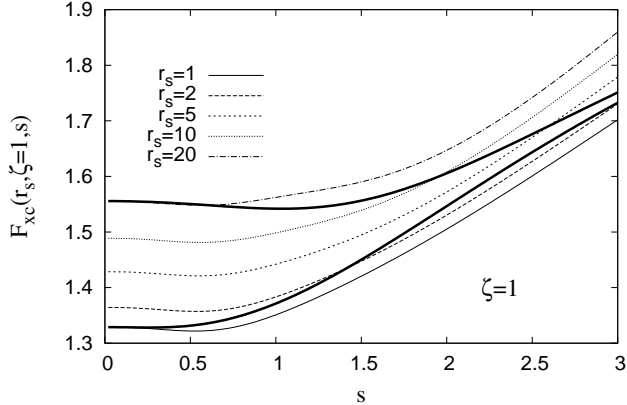}
\caption{ Enhancement factor $F_{xc}$ (see Eq. (\ref{e23}) for the fully-spin-polarized case 
($\zeta=1$), as a
function of the reduced gradient $s$ for several values of $r_s$ ($r_s=1$, 2, 5, 10, and
20). The thin lines represent the
ARPA+ enhancement factor whereas the thick lines are the
PBEsol enhancement factor for $r_s=1$ and $r_s=20$ respectively. The LSDA is 
$F_{xc}(r_s,\zeta=1,s=0)$.} 
\label{f7}
\end{figure}
Figs. \ref{f6} and \ref{f7} show the enhancement factor of ARPA+ compared to PBEsol as 
a function of the reduced gradient $s$, for several values of $r_s$, in the spin-unpolarized 
case and the fully-spin-polarized case, respectively. In both figures, the ARPA+ and PBEsol 
enhancement factors agree well at small gradients (for $s\leq 0.5$), but for $s>> 0.5$  
ARPA+ shows more exchange-correlation nonlocality than PBEsol.
%
\begin{figure}
\includegraphics[width=\columnwidth]{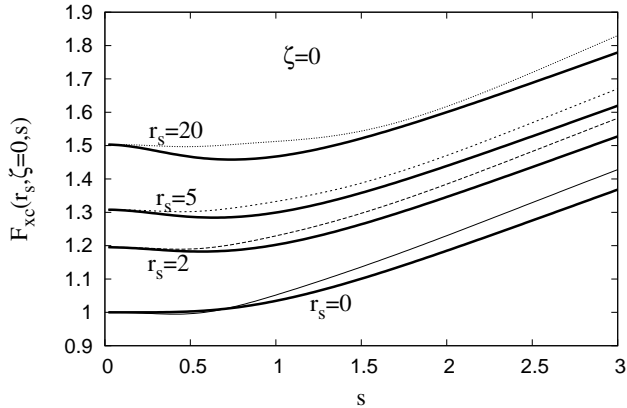}
\caption{ Comparison of $F^{ARPA+}_{xc}(r_s,\zeta=0,s)$ (shown with thin lines) and 
$F^{AM05}_{xc}(r_s,\zeta=0,s)$ 
(shown with thick lines) for
several values of $r_s$ ($r_s=0$, 2, 5, and 20).}
\label{f7bis}
\end{figure}
%
\begin{figure}
\includegraphics[width=\columnwidth]{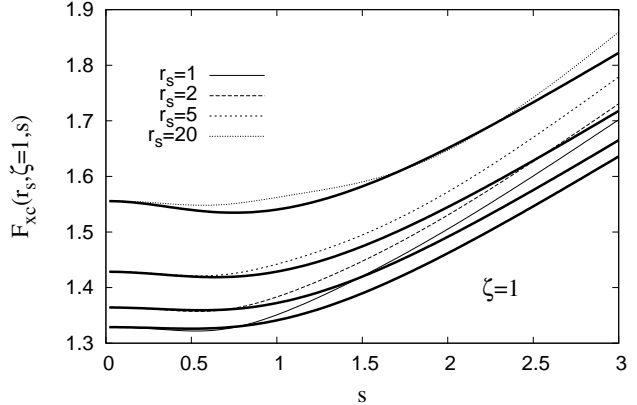}
\caption{ Comparison of $F^{ARPA+}_{xc}(r_s,\zeta=1,s)$ (shown with thin lines) and
$F^{AM05}_{xc}(r_s,\zeta=1,s)$ of Ref. \cite{AM05b}
(shown with thick lines) for
several values of $r_s$ ($r_s=1$, 2, 5, and 20).}
\label{f7bisb}
\end{figure}

Figs. \ref{f7bis} and \ref{f7bisb} show a comparison between the ARPA+ GGA and AM05 GGA 
enhancement factors, for 
the 
spin-unpolarized and fully spin-polarized cases. Up to $s=0.5$, $F^{ARPA+}_{xc}(r_s,\zeta,s)$ 
and 
$F^{AM05}_{xc}(r_s,\zeta,s)$ agree very well. For 
$s\geq 0.5$, $F^{ARPA+}_{xc}(r_s,\zeta,s)$ shows slightly more nonlocality than 
$F^{AM05}_{xc}(r_s,\zeta,s)$, and, even if 
this difference is small, 
it has noticeable effects for the lattice constants
of bulk solids.  Overall, our ARPA+ 
confirms the AM05 construction for correlation.

\section{Tests of the ARPA+ GGA xc energy functional}
\label{sec4}
\noindent

In this section we test
our functionals for jellium surfaces, atoms, molecules, and bulk solids. 
The calculations use the spin-scaling relation of Eq. (\ref{e16}). 

\subsection{Jellium surfaces}
\label{ss1}

\begin{figure}
\includegraphics[width=\columnwidth]{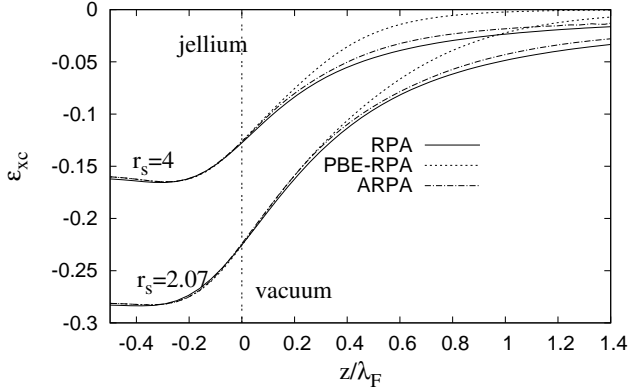}
\caption{
RPA exchange-correlation energy (hartree) per particle $\epsilon_{xc}$ at position $z$
versus $z/\lambda_F$, at surfaces of two jellium slabs.
The bulk parameters are $r_s=2.07$ and $r_s=4$. Both jellium slabs have the width
$d=3.2\lambda_F$. 
The edges of the positive background are at $z=0$.
The differences at large $z$, emphasized here by plotting
$\epsilon_{xc}$ instead of $n \epsilon_{xc}$, are not important for the surface energy.
}
\label{f8}
\end{figure}

In Fig. \ref{f8} we show $\epsilon_{xc}^{RPA}$ given by Eq. (\ref{e12}), $\epsilon_{xc}^{ARPA}$ 
given by Eq. (\ref{e21bis}), and $\epsilon_{xc}^{PBE-RPA}$ of Ref. \cite{YPK}, 
for two thick jellium slabs of bulk parameters $r_s=2.07$ and $r_s=4$. We use accurate LSDA 
orbitals and densities as in Refs. \cite{PE2,CPP2,PCP}. ARPA fits well the exact 
RPA until $s\approx 20$, showing that the Airy gas and the jellium surfaces are very close 
related, as expected.  
%

In Table \ref{table1} we report the ARPA and ARPA+ jellium surface exchange and xc energies. The
$\sigma^{ARPA+}_{x}$ are close to but worse than $\sigma^{LAG}_{x}$. The $\sigma^{ARPA}_{xc}$
are between $\sigma^{RPA}_{xc}$ and $\sigma^{PBE-RPA}_{xc}$ for
$r_s<\sim 3$, but lower than both others for $r_s>\sim 4$. 
The $\sigma^{ARPA+}_{xc}$ are reasonably close to $\sigma^{LSDA}_{xc}$
and $\sigma^{DMC}_{xc}$ (see Ref. \cite{WHFGG}), but are surprisingly lower and less accurate 
than 
$\sigma^{LSDA}_{xc}$.
\begin{table}[htbp]
\footnotesize
\caption{ Jellium surface exchange and exchange-correlation energies 
($\mathrm{erg/cm^{2}}$) for LSDA, PBE, and ARPA+ in and beyond the random phase approximation. 
We also show the jellium surface exchange and exchange-correlation energies beyond RPA, for 
LAG GGA, AM05 GGA, PBEsol GGA, and TPSS meta-GGA of Ref. \cite{TPSS}.
The 
exact values of $\sigma^{exact}_{x}$ and $\sigma^{RPA}_{xc}$ are from Ref. \cite{PE2}, and the 
fixed-node difussion Monte Carlo (DMC) $\sigma^{DMC}_{xc}$ values are interpolations and 
extrapolations of the estimates of Ref. \cite{WHFGG} (see Table II of Ref. \cite{CPDGP}). 
To interpolate or 
extrapolate $r_s$ we recommend Eq. (15) of Ref. \cite{APF}. 
($1\mathrm{hartree}/\mathrm{bohr}^2=1.557\times 10^6 \mathrm{erg}/\mathrm{cm}^{2}$.)  }
\begin{tabular}{|l|l|l|l|l|l|}
   \multicolumn{1}{c}{ } &
   \multicolumn{1}{c}{ } &
   \multicolumn{1}{c}{ } &
   \multicolumn{1}{c}{ } \\  \hline
$r_{s}$ & 2 & 3 & 4 & 6 \\  \hline
$\sigma^{LSDA}_{x}$ & 3036 & 669 & 222 & 43.6 \\  \hline
$\sigma^{PBE}_{x}$ & 2436 & 465 & 128 & 11.8 \\  \hline
$\sigma^{PBEsol}_{x}$ & 2666 & 540 & 162 & 22.9 \\  \hline
$\sigma^{TPSS}_{x}$ & 2553 & 498 & 141 & 15.4 \\  \hline
$\sigma^{LAG}_{x}$ & 2908 & 619 & 198 & 34.3 \\  \hline
$\sigma^{LAA}_{x}$ & 2896 & 615 & 196 & 33.6 \\  \hline
$\sigma^{AM05}_{x}$ & 2934 & 627 & 201 & 35.4 \\  \hline
$\sigma^{ARPA+}_{x}$ & 2941 & 626 & 199 & 34.6 \\  \hline
$\sigma^{exact}_{x}$ & 2624 & 526 & 157 & 22 \\  \hline\hline

$\sigma^{LSD-RPA}_{xc}$ & 3403 & 781 &269 & 56 \\  \hline
$\sigma^{PBE-RPA}_{xc}$ & 3318 & 760 &262 & 55 \\  \hline
$\sigma^{ARPA}_{xc}$ & 3366 & 764 & 260 & 53 \\  \hline
$\sigma^{RPA}_{xc}$ & 3467 & 801 & 278 & 58 \\  \hline\hline

$\sigma^{LSDA}_{xc}$ & 3354 & 764 &  261 & 53 \\  \hline
$\sigma^{PBE}_{xc}$ & 3265 & 741 & 252 & 52 \\  \hline
$\sigma^{PBEsol}_{xc}$ & 3374 & 774 & 267 & 56 \\  \hline
$\sigma^{TPSS}_{xc}$ & 3380 & 772 & 266 & 55 \\  \hline
$\sigma^{LAG}_{xc}$ & 3226 & 714 & 237 & 43.7 \\  \hline
$\sigma^{AM05}_{xc}$ & 3414 & 782 & 270 & 56.7 \\  \hline
$\sigma^{ARPA+}_{xc}$ & 3313 & 745 & 250 & 50 \\  \hline
$\sigma^{RPA+}_{xc}$ & 3413 & 781 & 268 & 54 \\  \hline
$\sigma^{DMC}_{xc}$ & 3392$\pm 50$ & 768$\pm 10$ & 261$\pm 8$ & 52.5$\pm$ ... \\  \hline\hline
\end{tabular}
\label{table1}
\end{table}

\subsection{Spherical atoms}
\label{ss2}

In Table \ref{table2} we calculate the ARPA+ exchange and correlation energies of several 
atoms and ions. We use spin-restricted analytic Hartree-Fock orbitals \cite{CR11} and 
densities. (The difference 
between Hartree-Fock orbitals and Kohn-Sham orbitals is small for atoms.)
For every atom and ion of Table \ref{table2}, ARPA+ GGA improves the LSDA results, but it is still a 
poor approximation in comparison with GGA's constructed for atoms and molecules, such as PBE GGA 
\cite{PBE,PCSB}.  
\begin{table}[htbp]
\footnotesize
\caption{ Exchange and correlation energies (in hartrees) of several spherical atoms and ions
with spin-restricted Hartree-Fock orbitals and densities \cite{CR11}. Exact correlation 
energies are from Ref. \cite{CGDPFF}. 
PBE GGA, not shown in the table, has the mean absolute errors (m.a.e.): 0.0476 for exchange and 
0.01563 for correlation. (See also Table V of Ref. \cite{PTSS}.) 
}
\begin{tabular}{|l|l|l|l|l|l|l|l|l|}
   \multicolumn{1}{c}{ } &
   \multicolumn{1}{c}{ } &
   \multicolumn{1}{c}{ } &
   \multicolumn{1}{c}{ } \\  \hline
 & $E^{LSDA}_x$ & $E^{ARPA+}_x$ & $E^{HF}_x$ & $E^{LSDA}_c$ & $E^{ARPA+}_c$ & $E^{exact}_c$ \\  
\hline
H              & -0.268  & -0.280 &-0.313 &  -0.0222 & -0.0199 & 0 \\
He             & -0.884  & -0.925 &-1.026 &  -0.1125 & -0.1030 & -0.0420\\
$\rm{Li}^+$    & -1.421  & -1.486 &-1.652 &  -0.1346 & -0.1233 & -0.0435\\
$\rm{Be}^{2+}$ & -1.957  & -2.047 &-2.277 &  -0.1504 & -0.1378 & -0.0443\\
Li             & -1.538  & -1.603 &-1.781 &  -0.1508 & -0.1378 & -0.0453\\
$\rm{Be}^+$    & -2.168  & -2.261 &-2.507 &  -0.1727 & -0.1578 & -0.0474\\
Be             & -2.312  & -2.408 &-2.667 &  -0.2240 & -0.2058 &-0.0943\\
$\rm{B}^+$     & -3.036  & -3.157 &-3.492 &  -0.2520 & -0.2317 &-0.1113\\
$\rm{Ne}^{6+}$ & -6.634  & -6.886 &-7.594 &  -0.3336 & -0.3069 &-0.1799\\
N              & -5.893  & -6.047 &-6.596 &  -0.4273 & -0.4016 &-0.1883\\
Ne             & -11.033 & -11.220&-12.109&  -0.7428 & -0.7084 &-0.3905\\
Ar             & -27.863 & -28.118&-30.190&  -1.4242 & -1.3723 &-0.7222\\ \hline
m.a.e.         & 0.600   & 0.481  &       &   0.1865 & 0.1664  &        \\ \hline
\end{tabular}
\label{table2}
\end{table}
\begin{table}[htbp]
\footnotesize
\caption{ Change in xc energy (hartree) of an atom due to removal of a shell of valence 
electrons($\Delta 
E_{xc}= E_{xc}^{atom} - E_{xc}^{ion}$). The 
calculation is based on the exchange and correlation energies listed in Table \ref{table2} of this 
work and in 
Table VI of Ref. \cite{PTSS}. }
\begin{tabular}{|l|l|l|l|l|l|}
   \multicolumn{1}{c}{ } &
   \multicolumn{1}{c}{ } &
   \multicolumn{1}{c}{ } &
   \multicolumn{1}{c}{ } \\  \hline
 & $\Delta E_{xc}^{LSDA}$ & $\Delta E_{xc}^{ARPA+}$ & $\Delta E_{xc}^{PBE}$ &$\Delta E_{xc}^{exact}$ 
\\  \hline
Li $\rightarrow \rm{Li}^+$ & -0.133 & -0.132 & -0.138  & -0.131 \\
Be $\rightarrow \rm{Be}^{+2}$ & -0.429 & -0.430 & -0.438 & -0.440 \\
Ne $\rightarrow \rm{Ne}^{+6}$ & -4.808 & -4.737 & -4.793 & -4.726 \\ \hline
\end{tabular}
\label{table3}
\end{table}

In Table III we show the xc contribution to the valence-shell removal energy (a quantity that can be 
accurately measured experimentally \cite{PTSS}) of three atoms (Li, Be, and Ne). We observe that the 
ARPA+ systematically improves the LSDA results, competing in accuracy with the PBE GGA.

\subsection{ Atomization energies of molecules}
\label{ss3}

The AE6 test set \cite{LT} of atomization energies of molecules has only six molecules ($\rm{SiH}_4$, 
SiO, $\rm{S}_2$, $\rm{C}_3\rm{H}_4$, $\rm{C}_2\rm{H}_2\rm{O}_2$, and $\rm{C}_4\rm{H}_8$) and was 
constructed to reproduce 
the errors of density functionals for larger molecular sets, providing a quick but
representative evaluation of the accuracy of density functionals for molecules.
In Table \ref{table4} we show the errors (in kcal/mol) of the AE6 atomization energies for ARPA+ GGA, 
ARPA GGA, PBE GGA, PBEsol GGA, and AM05 GGA. The errors given by ARPA+ GGA and ARPA GGA are 
practically the 
same, in accord with the work of Ref. \cite{YPK}, and show that the RPA+ short-range 
correction does not have an important effect on the atomization energies of molecules. 
Although
our GGA short-range correction to RPA is important for total energies, it tends to cancel out of
energy differences for processes in which the electron number remains unchanged (as in Tables 
\ref{table1} 
and 
\ref{table4} but
not Tables \ref{table2} and \ref{table3}).
The accuracy 
of 
the ARPA+ for the AE6 test 
is close to that
of PBEsol, with both reducing the LSDA error by
by more than a factor of two.
\begin{table}[htbp]
\footnotesize
\caption{ The errors (kcal/mole) of the atomization energies of the
AE6 set of molecules. We use the $6-311+G(3df,2p)$ basis set in the Gaussian03 code.
The AM05 atomization energies of the AE6 set of molecules were calculated
in Ref. \cite{PRCCS}, using the spin-polarized version of AM05 given in Ref. \cite{AM05b}. 
The LSDA mean error (ME) is 77.3 kcal/mole and its mean absolute 
error (MAE) is 77.3 kcal/mole \cite{PRCVSCZB}. The TPSS meta-GGA of Ref. \cite{TPSS} gives 
ME=4.2 
kcal/mole, and MAE=6.0 kcal/mole. 
The AE6 mean atomization energy is 517 kcal/mole.
(1 hartree = 627.5 kcal/mole.) (For ARPA+ and ARPA, we used PBEsol densities.) }
\begin{tabular}{|l|l|l|l|l|l|}
   \multicolumn{1}{c}{ } &
   \multicolumn{1}{c}{ } &
   \multicolumn{1}{c}{ } &
   \multicolumn{1}{c}{ } \\  \hline
 & PBE & ARPA+ & ARPA & PBEsol & AM05 \\  \hline
$\rm{SiH}_4$ & -9.2 & 10.1 & 9.9 & 1.3 & 7.6\\
SiO & 3.6 & 11.2 & 12.3 & 12.9 & 13.5 \\
$\rm{S}_2$ & 13.1 & 18.4 & 19.2 & 21.9 & 21.6 \\
$\rm{C}_3\rm{H}_4$ & 16.4 & 46.0 & 50.6 & 45.1 & 48.1 \\
$\rm{C}_2\rm{H}_2\rm{O}_2$ & 31.8 & 60.1 & 65.7 & 64.7 & 66.6 \\
$\rm{C}_4\rm{H}_8$ & 18.7 & 70.6 & 78.7 & 69.6 & 75.0 \\  \hline
ME & 12.4 & 36.1 & 39.4 & 35.9 & 38.7 \\
MAE & 15.5 & 36.1 & 39.4 & 35.9 & 38.7 \\  \hline\hline
\end{tabular}
\label{table4}
\end{table}

While our ARPA overbinds molecules (and this overbinding is only slightly reduced in ARPA+),
the full RPA apparently underbinds molecules \cite{Furche}. Thus, even at the RPA level, the Airy
gas xc energy density does not seem to transfer very accurately to molecules: much better atomization
energies are predicted by standard functionals like the PBE GGA \cite{PBE} or the TPSS meta-GGA 
\cite{TPSS}.
GGA overbinding
of molecules typically goes together with GGA underestimation of the magnitude of the 
exchange-correlation energy of an atom, which we found for LSDA and ARPA+ but not so much for PBE in 
Table \ref{table2}.

\subsection{ Equilibrium lattice constants of solids}
\label{ss4}

In Table \ref{table5} we test the ARPA+ GGA for a simple metal (Na), a semiconductor (Si), a 
transition metal (Cu), and an ionic solid (NaCl). 
The ARPA+ GGA lattice constants are longer than the PBEsol ones, but shorter than the PBE values,
except for NaCl where ARPA+ is close to PBE.
These trends are plausible from the enhancement factors plotted in Figs. \ref{f6} and \ref{f7},
and the maximum $s$ values reported in Ref. \cite{CPRPLPVA}.
These calculations also suggest that the correct second-order gradient expansion for exchange 
\cite{AK}, 
employed in the construction of the PBEsol GGA, is the most promising path 
toward an accurate and nonempirical GGA for 
solids.

\begin{table}[htbp]
\footnotesize
\caption{Lattice constants (in \AA) calculated with the Gaussian03 code as in Ref. 
\cite{PRCVSCZB} and
compared to experimental values
corrected to the static-lattice limit \cite{PRCVSCZB,APBAF}. (For ARPA+, we used PBEsol 
densities.)
}
\begin{tabular}{|l|l|l|l|l|l|}
   \multicolumn{1}{c}{ } &
   \multicolumn{1}{c}{ } &
   \multicolumn{1}{c}{ } &
   \multicolumn{1}{c}{ } \\  \hline
\rm{Solid} & LSDA & PBE & PBEsol & ARPA+ & Exper.\\  \hline
Na & 4.049 & 4.199 & 4.159 &  4.207 & 4.210 \\
Si & 5.410 & 5.479 & 5.442 &  5.470 & 5.423 \\
Cu & 3.530 & 3.635 & 3.578 &  3.605 & 3.596 \\
NaCl & 5.471 & 5.696 & 5.611 &5.716 & 5.580 \\ \hline
ME & -0.087 & 0.050 & -0.005 & 0.045 &  \\ 
MAE & 0.087 & 0.056 & 0.030 & 0.049 & \\ \hline \hline
\end{tabular}
\label{table5}
\end{table}

The Gaussian03 code that we use gives lattice constants that are on average a little too
long \cite{CPRPLPVA}.  The LSDA lattice constants calculated with the more-accurate WIEN2K code
are \cite{HTB}: Na 4.047, Si 5.407, Cu 3.522, and
NaCl 5.465. Thus, extensive and more accurate lattice constants calculations
need to be performed for our ARPA+.

\section{ Conclusions}
\label{sec5}

In this paper we construct the RPA correlation energy density of the Airy gas, using an accurate Airy 
gas 
model that has only 19 occupied orbitals. This approch can be generalized to other physical systems, 
such as 
a more sophisticated edge electron gas that can include curvature corrections (arising from 
nonlinearity of $v_{eff}(z)$).

We have constructed the ARPA GGA that accurately fits the RPA xc
energy density of the Airy gas, and we have corrected its 
short-range part in the framework of the RPA+ \cite{YPK} approach, developing the ARPA+ GGA 
entirely without empiricism. Because of the delocalization of the electrons in the Airy gas, 
our ARPA+ GGA has nearly the correct Airy-gas correlation energy.
Via our Figs. 8 and 9, our ARPA+ confirms the AM05 hypothesis
\cite{AM05} for the 
correlation functional compatible
with Airy-gas GGA exchange \cite{Vi2,AM05}.

By testing the  ARPA+ GGA for jellium surfaces, atoms, molecules, and bulk solids, we observe that  
the xc energy density of the Airy gas can be transferred successfully
to a very similar system such as the jellium surface, but less successfully to a very different
system like
a bulk solid, an atom, or a molecule.
However, the ARPA+ GGA mildly improves the LSDA results for lattice constants and atomization
energies, without much worsening the already-good surface exchange-correlation energies.
        
We would have liked to replace the RPA+ method by the more sophisticated inhomogeneous
Singwi-Tosi-Land-Sj\H{o}lander (ISTLS) \cite{DWG,CPDGP}, but were not able to achieve sufficiently 
accurate 
numerical results
for the correlation energy densities thereof.  
The future use of ISTLS could refine
our input, and provide an energy density (not just an integrated energy) for the 
short-range correction to RPA.
Other possible future refinements could include the use of different
reference systems for the bulk and surface of a solid \cite{AM05,AM11}, replacing the Airy gas
by a more sophisticated
example of the edge electron gas, or replacing the GGA functional 
form by the meta-GGA \cite{TPSS}.
We suspect \cite{PRCVSCZB2,PRCCS} that the meta-GGA form is needed to achieve simultaneous high 
accuracy 
for atoms, molecules, and solids near equilibrium. In fact the TPSS meta-GGA \cite{TPSS,TSSP11} is 
already 
close to being such a general-purpose semilocal functional,
and a revised TPSS \cite{PRCCS} with improved lattice constants may be even closer.

We note however that there are two formally unsatisfactory aspects of using the 
exchange-correlation
energy density of a nonuniform system as a reference for the construction of density functionals: (1) 
Except in the uniform electron gas, the energy density is neither
observable nor unique, since any function integrating to zero can be added 
to it
with no physical consequence.  Here, as in Refs. \cite{Vi2},\cite{AM05},\cite{PCR},\cite{CC11}, 
and 
\cite{CC111}, 
we have 
chosen the conventional \cite{TSSP11}
gauge for the energy density, but other choices should be explored.  (2) While the integrated 
exchange energy for a slowly-varying density 
is expressible in terms of the GGA ingredients $n$ and $\nabla n$, the conventional exchange energy 
density in this limit is not
so expressible, having a Laplacian term $\nabla^2 n^{2/3}$
which integrates to zero but has a divergent coefficient \cite{PW11,AM11}.  
As a result, the Airy-gas GGA
cannot predict accurate exchange energies for slowly-varying electron densities (e.g.,the jellium 
surface exchange energy), while  more standardly-constructed GGA's like PBEsol can do so 
\cite{PRCVSCZB} (our Table I).  
The Airy-gas GGA can at best work for the jellium surface by error cancellation between
exchange and correlation, which is possible for typical valence-electron densities but not under 
uniform density
scaling to the high-density limit where exchange dominates.

The GGA constructed here has no clear practical advantage over already-published ones.  Our
purpose is not to advocate its use, but to show what is obtained from the Airy-gas 
reference
system within a consistent implementation for correlation as well as exchange.

\begin{acknowledgment}
We thank Levente Vitos for providing us with the exact exchange energy per particle of the Airy gas, 
shown in Fig. \ref{f3} as the curve labelled "Airy gas".
We thank Ann Mattsson for comments on the manuscript.
L.A.C. thanks J.M. Pitarke for many valuable discussions and suggestions.
L.A.C. and J.P.P. acknowledge NSF support (Grant No. DMR05-01588).
\end{acknowledgment}


\begin{thebibliography}{[Vo]}
\bibitem{noteVE}
A linear effective potential does not imply a linear
external potential. A quantum Monte Carlo
calculation of the correlation energy density would have to start from the external potential.
\bibitem{noteG}
 We use a grid of 350 points nonuniformly distributed on the $z$-axis.
Our grid gives accurate RPA correlation energy densities for the region
of interest $s\geq 0.3$. Calculations with 70 and 90 unoccupied orbitals,
and with 450 grid points, gives unchanged RPA correlation energy
densities for $s\geq 0.3$. The biggest error of our numerical calculation
is in the region $s\leq 0.3$, where the Thomas-Fermi approximation becomes accurate.
\bibitem{noteSc}
The numerical RPA conventional correlation energy
density depends significantly on the slope value $F$ when they are
plotted against the standard reduced gradient for the correlation
energy $t=|\nabla n|/[4(3/\pi)^{1/6}n^{7/6}]$. Thus the Airy gas correlation
energy density is better described by $s_c$ of Eq. (\ref{e20}), or even
by $s$ of Eq. (\ref{e11}) (as proposed in AM05 GGA), than the standard reduced 
gradient 
for correlation $t$ that was derived in the linear response of the 
uniform electron gas.
\end{thebibliography}
\end{document}